\documentstyle[ltwol,epsfig]{article}

\def\Journal#1#2#3#4{{#1} {\bf #2}, #3 (#4)}

\def\NPB{{\em Nucl. Phys.} B}
\def\PLB{{\em Phys. Lett.}  B}

\def\ZPC{{\em Z. Phys.} C}
\def\JPL{\em JETP Lett.}

\def\be{\begin{equation}}
\def\ee{\end{equation}}
\def\bea{\begin{eqnarray}}
\def\eea{\end{eqnarray}}

\begin{document}

\title{GAUGE INVARIANT FIELD STRENGTH CORRELATORS IN QCD.}

\author{presented by A. DI GIACOMO}

\address{Dip. Fisica Universit\`a and INFN, Via Buonarroti 2 ed.B 56126
PISA, ITALY\\E-mail: digiacomo@pi.infn.it}

\author{M. D'ELIA, H. PANAGOPOULOS}

\address{Department of Natural Sciences University of Cyprus, PO Box
537 Nicosia, CY-1678, Cyprus\\E-mail: delia@dirac.ns.ucy.ac.cy\\
E-mail: haris@earth.ns.ucy.ac.cy}

\author{E. MEGGIOLARO}
\address{
Institut f\"ur Theoretische Physik Universit\"at Heidelberg
Philosophenweg 16 D-69120 Heidelberg Germany\\
E-mail: e.meggiolaro@sns.it}


\twocolumn[\maketitle\abstracts{
Gauge invariant correlators in QCD are studied on the lattice. A systematic
determination of the correlation lengths for gluon field strength
correlators and quark correlators is made. The measurement of the gluon and
quark condensates is discussed.
}]
\section{Introduction}
The gauge invariant gluon field strength correlators are defined as
\begin{equation}
{\cal D}_{\mu\nu\rho\sigma}(x,C) =
\langle 0 | T\left( G^a_{\mu\nu}(x) S^{ab}_{adj,C}(x,0) G^b_{\rho\sigma}(0)
\right) |0\rangle \label{eq:1}\end{equation}
with $S^{ab}_{adj,C}(x,0)$ the parallel transport from 0 to $x$ along the path
$C$, in the adjoint representation
\begin{equation}
S_{adj,C}(x,0) =
P\exp\left( i\,\int_{0,C}^x A_\mu^a(y) T^a_{adj}\,d y^\mu\right)
\label{eq:2}\end{equation}
$T^a_{adj}$ are the group generators in the adjoint representation.

${\cal D}_{\mu\nu\rho\sigma}(x,C)$ depends on the choice of $C$: in what
follows
we will take for $C$ a straight line, and drop the dependence on $C$.

Higher correlators are usually defined by parallel transport to a fixed
point~$x_0$.

Fermion correlators are defined as
\begin{equation}
S_i(x) =
\langle 0 | T \left( \bar\psi(x) S_{fund}(x,0) M^i \psi(0)\right) |0\rangle
\label{eq:3}\end{equation}
$M^i$ is a generic element of the Clifford algebra of the $\gamma$ matrices.
$S_{fund}$ is the analog of the transport (\ref{eq:2}) in the fundamental
representation, and again we have once and for all assumed for the path $C$ a
straight line.

By use of general covariance arguments\cite{1,2} ${\cal D}_{\mu\nu\rho\sigma}$
can be parametrized in terms of two independent invariant form factors
${\cal D}(x^2)$ and ${\cal D}_1(x^2)$
\begin{eqnarray}
{\cal D}_{\mu\nu\rho\sigma}(x) &=&
(g_{\mu\rho} g_{\nu\sigma} - g_{\mu\sigma} g_{\nu\rho})
\left[ {\cal D}(x^2) + {\cal D}_1(x^2)\right] +
\label{eq:4}\\
&& \hskip-30pt+(x_\mu x_\rho g_{\nu\sigma} - x_\mu x_\sigma g_{\nu\rho} -
x_\rho x_\nu g_{\mu\sigma} + x_\nu x_\sigma g_{\mu\rho})
\frac{\partial {\cal D}_1}{\partial x^2} \nonumber
\end{eqnarray}
Similarly one can prove that all the correlators (\ref{eq:3}) vanish by $T$,
$P$ invariance, except the correlator $S_0$ corresponding to $M^i = I$, the
identity matrix
\begin{equation}
S_0(x) =
\langle 0 | T \left( \bar\psi(x) S_{fund}(x,0) \psi(0)\right) |0\rangle
\label{eq:5}\end{equation}
and the vector correlator ($M^i = \gamma^\mu$), with $\gamma^\mu$ in the
direction of $x$
\begin{equation}
\frac{x^\mu}{|x|} S_V(x) =
\langle 0 | T \left( \bar\psi(x) S_{fund}(x,0) \gamma^\mu
\psi(0)\right) |0\rangle
\label{eq:6}\end{equation}
The physical interest of the above correlators stems from the basic idea of
the
ITEP sum rules\cite{3}: the long distance modes of QCD are described by a
slowly varying background, made e.g. of instantons, on which high momentum
perturbative fluctuations are superimposed. In  the O.P.E. low modes generate
the condensates, and high frequency modes the corresponding coefficient
functions
$C_n(x)$
\begin{eqnarray}
T\left( j_\mu(x) j_\nu(0)\right) &\simeq&
\sum_n C_n(x) O_n \label{eq:7}\\
&=& C_I(x) I + C_G(x) G_{\mu\nu}(0) G_{\mu\nu}(0) +
\nonumber\\
&+&C_\psi(x) \sum m_f \bar\psi_f(0)\psi_f(0) +\ldots
\nonumber
\end{eqnarray}
As is well known expressing the left hand side in terms of a dispersive
integral, relates masses and widths of resonances in $e^+ e^-\to hadrons$ to
the condensates
\[ G_2 = \frac{\beta(g)}{g}\langle 0| G^a_{\mu\nu}(0) G^a_{\mu\nu}(0)
|0\rangle
\quad,\quad
\langle 0|m_f \bar\psi_f(0)\psi_f(0)|0\rangle\]
(SVZ sum rules).

By use of this idea it was proposed in ref.\cite{4,5} that the gluon
condensate
$G_2$ could be determined from the spectrum of bound states of heavy $Q\bar Q$
systems. If the correlation length of the slow varying field, $\lambda$, is
much
bigger than the typical time of the bound system, then its effect is
in all respects a static Stark effect on the levels, and the gluon condensate
can be extracted from it.

A more detailed analysis\cite{6} involves ${\cal D}_{\mu\nu\rho\sigma}(x)$,
and
gives a shift depending on the parameter
\begin{equation}
\rho = \lambda \frac{m_q \alpha_s^2}{4} \label{eq:8}\end{equation}
where $4/m_q \alpha_s^2$ is the typical time of the low lying levels
of the
system, and $\lambda$ is the correlation length defined as
\begin{equation}
{\cal D}(x) \mathop\simeq_{|x|\to\infty} G_2 \exp(-x/\lambda)
\label{eq:9}\end{equation}
Measuring $\lambda$ was the motivation to investigate for the first time
${\cal D}_{\mu\nu\rho\sigma}(x)$ on the lattice\cite{7}. The computation was
done in quenched $SU(2)$ and gave a surprisingly small value of $\lambda$
\begin{equation}
\lambda \simeq 0.16 \,{\rm fm} \label{eq:10}\end{equation}
shaking the very bases of the SVZ approach.

A stochastic model of the vacuum was subsequently developed\cite{1,2}, in
which
observables are expressed in terms of invariant field strength correlators,
and
a cluster expansion is made. The basic assumption of the model is that higher
order clusters are negligible.

The quark correlator $S_0$ instead, known also as ``non local fermion
condensate'' enters in the construction of the wave functions of hadrons, in
particular in the computation of the pion form factor\cite{8}.

A technical breakthrough, the use of cooling or smearing procedures to polish
short distance fluctuations, leaving long distance physics unchanged\cite{9},
allowed a better determination of correlators\cite{10,11,12}.

The physical motivations to study correlators on the lattice are in conclusion
\begin{itemize}
\item[(i)]  understanding $\lambda$, and the basis of SVZ sum rules.
\item[(ii)] measuring condensates from first principles.
\item[(iii)] more generally providing inputs from first principles to
the community of stochastic vacuum practitioners.
\end{itemize}
\section{Cooling-smearing correlators}
Short range fluctuations in lattice configurations can be smoothed off by a
local cooling procedure which consists in replacing a link by the sum of the
inverse ``staples'' attached to it
\begin{equation}
\hbox{
\begin{minipage}{0.8\linewidth}
\vskip-0.1in
\epsfxsize0.9\linewidth
\epsfbox{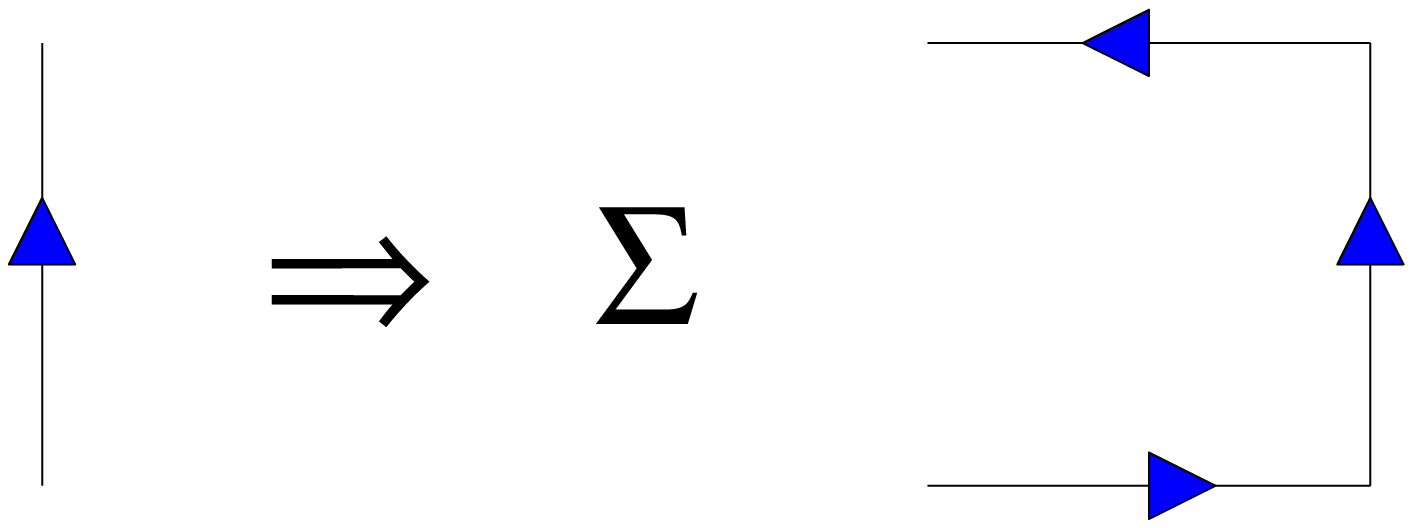}
\end{minipage}
}
\label{eq:10x}\end{equation}
Since the action density is
\begin{equation}
S\sim \sum_{\mu\nu}\left(1 - \frac{1}{N_c} Tr\Pi_{\mu\nu}\right)\quad
{\rm with}\; \Pi_{\mu\nu} =
\hbox{
\begin{minipage}{0.18\linewidth}
\epsfxsize0.5\linewidth
\epsfbox{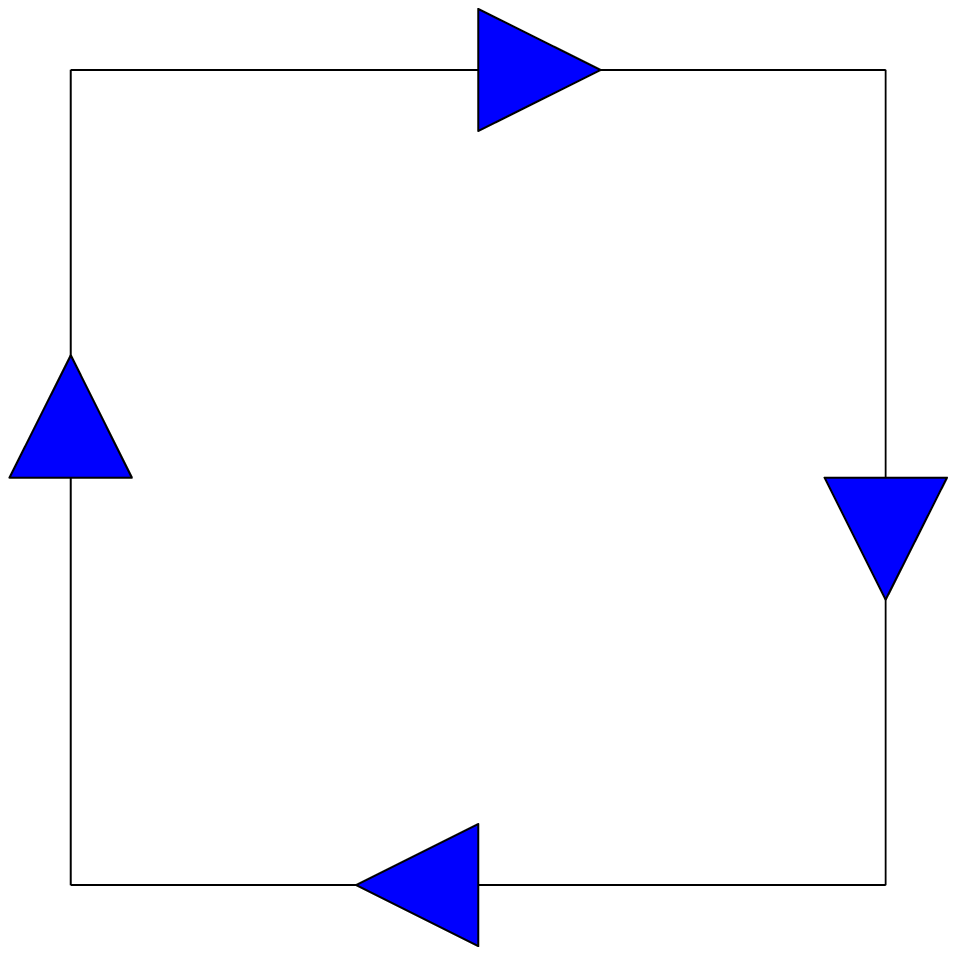}
\end{minipage}
}
\label{eq:1x}\end{equation}
this procedure makes locally $S=0$. In the euclidean region $S$ plays the
role of energy and the replacement (\ref{eq:10x}) locally minimizes $S$,
whence the terminology ``cooling''.

Like any local procedure cooling $n_t$ times affects distances $d$ by a
diffusion process, with
\begin{equation}
d^2 \sim n_t \label{eq:11}\end{equation}
According to eq.(\ref{eq:11}), $n_t$ can be made sufficiently large to
eliminate short range fluctuations but not enough to modify
long range correlations\cite{9}. Fluctuations are thus reduced by orders of
magnitude without changing long distance physics.

Gauge invariant correlators can be represented on lattice by the following
operators\cite{7}
\par\noindent
\begin{minipage}{0.6\linewidth}
${\cal D}^L_{\mu\nu\rho\sigma}=\left\langle
\hbox{
\epsfxsize0.4\linewidth
\epsfbox{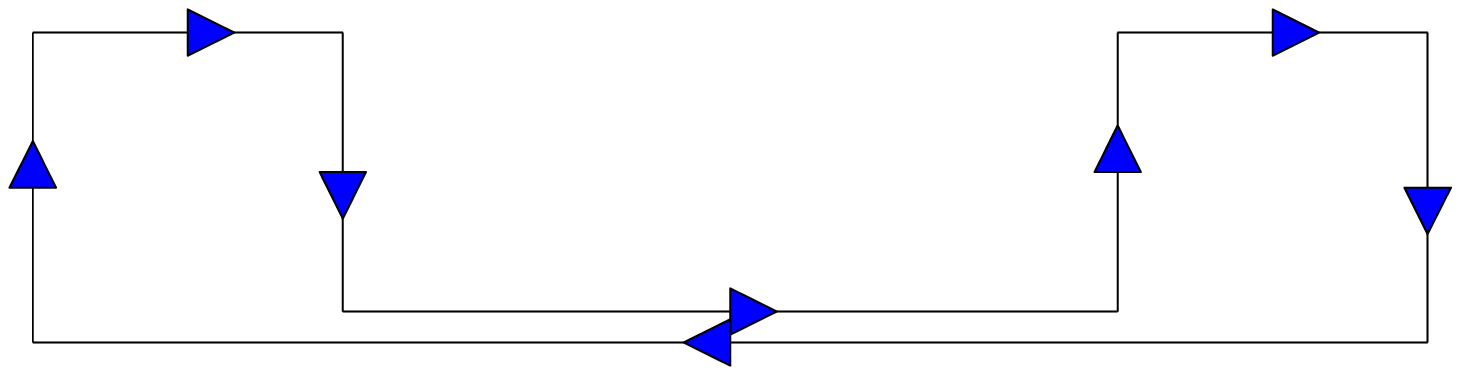}
}
\right.
$
\end{minipage}
\begin{minipage}{0.39\linewidth}\hskip-30pt
$-\frac{1}{N_c}\left.
\hbox{
\epsfxsize0.5\linewidth
\epsfbox{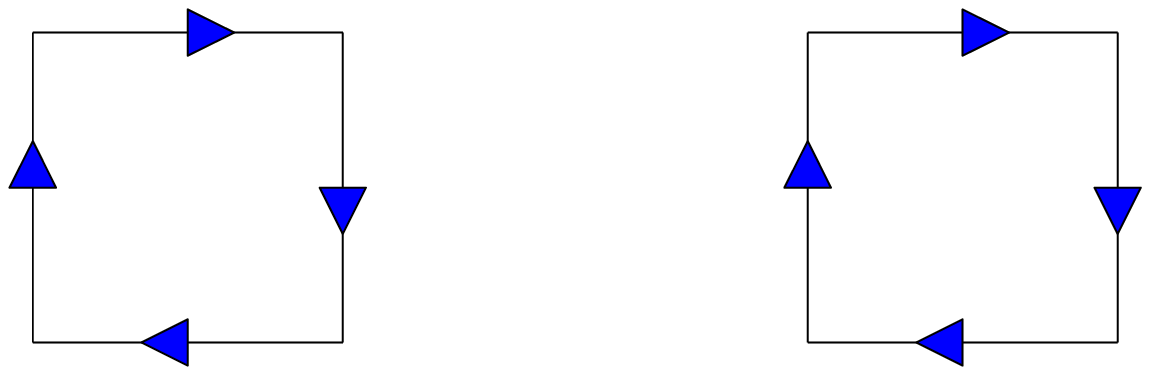}}\right\rangle$
\end{minipage}\par\noindent
\begin{minipage}{\linewidth}
\vskip-10pt\hskip52pt$\Pi_{\mu\nu}$
\hskip26pt$\Pi_{\rho\sigma}$
\hskip30pt$\Pi_{\mu\nu}$
\hskip10pt$\Pi_{\rho\sigma}$
\end{minipage}
A series expansion in $a$ gives
\begin{equation}
{\cal D}^L_{\mu\nu\rho\sigma}(d) \simeq
Z^2 a^4 {\cal D}_{\mu\nu\rho\sigma}(d) + {\cal
O}(a^6)\label{eq:12}\end{equation}
By the cooling procedure ${\cal O}(a^6)$ terms disappear and possible
renormalizations $Z$ of the field $G_{\mu\nu}$ tend to 1. So finally
\begin{equation}
{\cal D}^L_{\mu\nu\rho\sigma}(d) \simeq
 a^4 {\cal D}_{\mu\nu\rho\sigma}(d a)\label{eq:13}\end{equation}
In order to use cooling profitably the distance $d$ has to be $\sim 3-4$
lattice
spacings at least. At a given $\beta$ this corresponds to some physical length
$l_{min}$. If we want $l_{min}$ to be small, say 0.1~fm, since the lattice
must
be at least 1~fm across, the lattice size must be $(10 l_{min})^4$. Going to
small distances requires big lattices.
\section{Correlators, OPE and renormalons}
The SVZ sum rules are based on the OPE of the correlator
\begin{eqnarray}
\Pi_{\mu\nu}(q) &=& \int d^4x e^{i q x}
\langle 0 | T\left( j_\mu(x) j_\nu(0)\right) |0\rangle
\label{eq:14}\\
&=& \Pi(q^2) \left(g_{\mu\nu} q^2 - q_\mu q_\nu\right)\nonumber\end{eqnarray}
The OPE gives
\begin{equation}
\Pi(q^2) \simeq c_1 I + c_2 \frac{G_2}{q^4} +
\ldots\label{eq:15}\end{equation}
The first term corresponds to the perturbative expansion. That expansion,
however, is not Borel summable, and is ambiguous by terms of
order
$\mu^4/q^4$, with $\mu$ the renormalization scale (Renormalons).
As a consequence the second term in eq.(\ref{eq:15}) is intrinsically
undefined.
This is a basic and unavoidable drawback of perturbation theory, reflecting
the
fact that perturbative vacuum is not the ground state\cite{13}.

However, keeping
the first few terms in the perturbative expansion of
$c_1$ and $c_2$ gives a consistent phenomenology\cite{3}.

The same happens for the correlators. The OPE of the invariant form factors
has the form
\begin{equation}
{\cal D}(x^2) \mathop\simeq_{x^2\to 0} \frac{c_1}{x^4} + c_2 G_2 + {\cal
O}(x^2)
\label{eq:16}\end{equation}
The second term is again undetermined by renormalons coming from the
perturbative expansion of $c_1$. As in the SVZ sum rules we shall assume that
the
first few terms of the perturbative
expansion work, and use it to determine $G_2$. We shall parametrize the
lattice determination of ${\cal D}(x^2)$ and ${\cal D}^{(1)}(x^2)$ as
\begin{eqnarray}
\frac{1}{a^4} {\cal D}_L(x^2) &=&
\frac{a}{|x|^4}e^{-|x|/\lambda_a} + A_0 e^{-|x|/\lambda}
\label{eq:17a}\\
\frac{1}{a^4} {\cal D}^{(1)}_L(x^2) &=&
\frac{a_1}{|x|^4}e^{-|x|/\lambda_a} + A_1 e^{-|x|/\lambda}
\label{eq:17b}\end{eqnarray}
Eq.(\ref{eq:17a}), (\ref{eq:17b}) obey the OPE eq.(\ref{eq:16}), and reflect
the existence of a mass gap in the theory.

In ref.\cite{7} the first term of
eq.(\ref{eq:16}), (\ref{eq:17a}), (\ref{eq:17b})
 was computed in perturbation
theory and subtracted. The residual term was an exponential and
$\lambda$ could be extracted from it, giving $\lambda = 0.16$~fm.

In
ref.\cite{10,11} quenched $SU(3)$ was studied. The above parametrization
gave a
good fit to the data (fig.1)  with $\lambda = 0.22$~fm and $G_2 = (0.14\pm
0.02)$~GeV$^4$, a value
larger by an order of magnitude than the phenomenological value\cite{3}. In
ref\cite{12} full QCD with 4 staggered fermions was studied, at quark
masses $a m_q
= 0.01$, $am_q = 0.02$. The results are in this case\cite{12}
\par\noindent
\begin{minipage}{0.9\linewidth}
\epsfxsize0.9\linewidth
{
\epsfbox{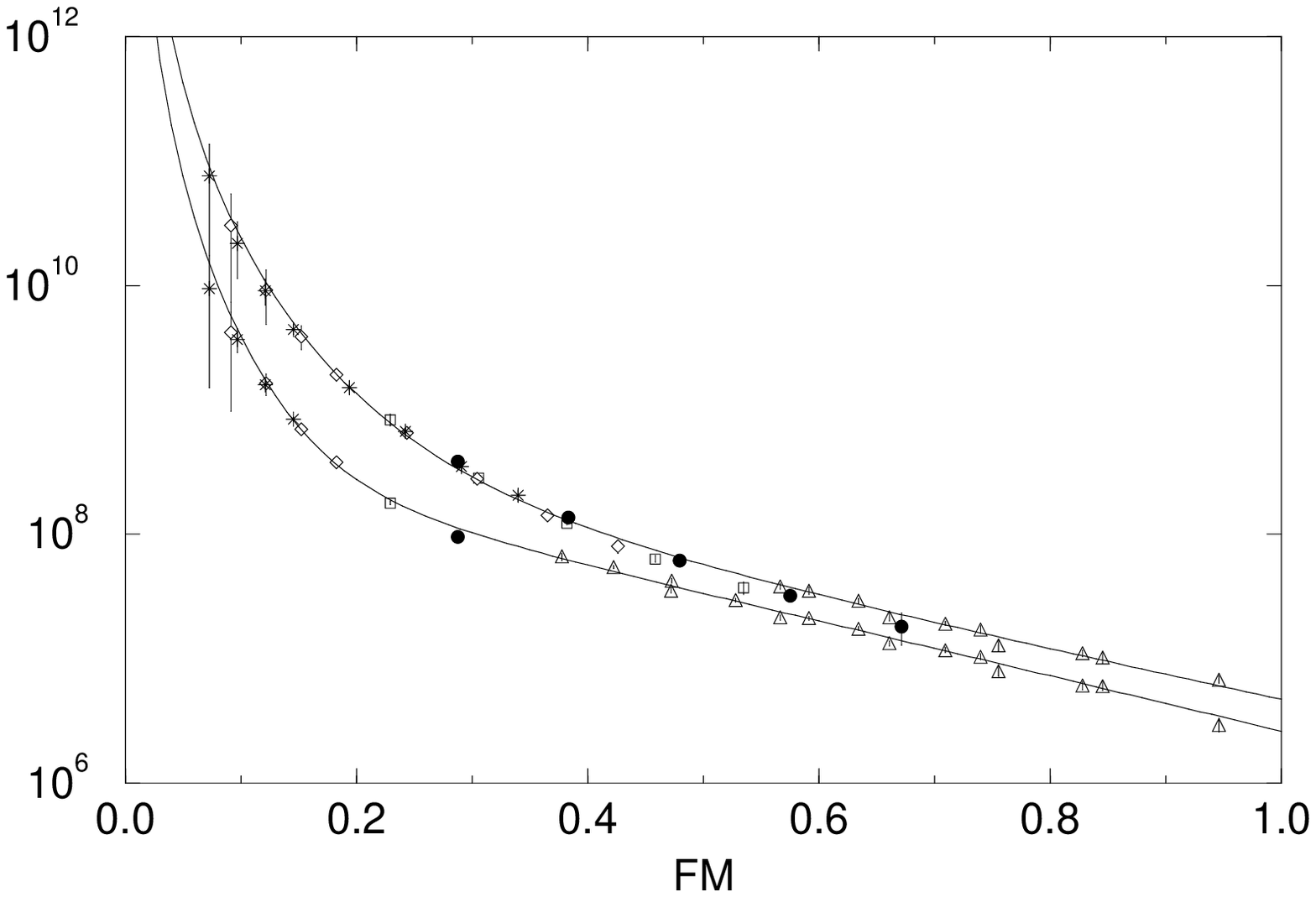}}
{\bf Fig.1}
$D_{||}^L/a^4 = {\cal D} + {\cal D}_1 + x^2 \partial {\cal D}_1/\partial x^2$
and $D_\perp/a^4 = {\cal D}+ {\cal D}_1$ versus $x$. The lines
correspond to the best fit to eq.(19),(20).
\end{minipage}
\begin{eqnarray*}
a m_q = .01&& \lambda = (.34\pm.02)\,{\rm fm}\quad
G_2 = .015 \pm .003\,{\rm GeV}^4 \label{eq:18a}\\
a m_q = .02&& \lambda = (.29\pm.02)\,{\rm fm}\quad
G_2 = .031 \pm .005\,{\rm GeV}^4 \label{eq:18b}
\end{eqnarray*}
A common feature to all the determinations is that $|A_1|\simeq A_0/10$.
In full QCD the correlation length is  bigger, and could agree with the
basic philosophy of ref.\cite{3}. Also the value of the condensate $G_2$ is
smaller than the quenched value and agrees with phenomenology.

By use of the relation\cite{14}
\[ \frac{d}{d m_f} G_2 = - \frac{24}{b_0} \langle \bar\psi\psi\rangle\qquad
b_0 = 11 - \frac{2}{3} N_f\]
one can extrapolate in $m_f$ to the physical value of $G_2$ getting
\begin{equation}
G_2 \simeq 0.022 \pm.006\,\,{\rm GeV}^4 \label{eq:19}\end{equation}
in agreement with sum rules determination\cite{15}.

Similar arguments allow to extract $G_2$ from the measurement of the average
value of the density of action (plaquette): again the level of rigour is the
same as for SVZ sum rules\cite{3}, at least if only a few terms are kept of
the
perturbative expansion of the coefficients in the OPE\cite{16,17,18}.

A detailed analysis of the behaviour of the correlators at finite
temperature, in the vicinity of the deconfining transition $T_c$ was
made in ref.\cite{11}. There $O(4)$ invariance is lost reducing to $O(3)$
and 5
independent form factors exist.

The main result is that magnetic correlators are unchanged across $T_c$, while
electric correlators have a sharp drop (fig.2,fig.3).
\par\noindent
\begin{minipage}{0.9\linewidth}
\epsfxsize0.8\linewidth
\rotatebox{270}{
\epsfbox{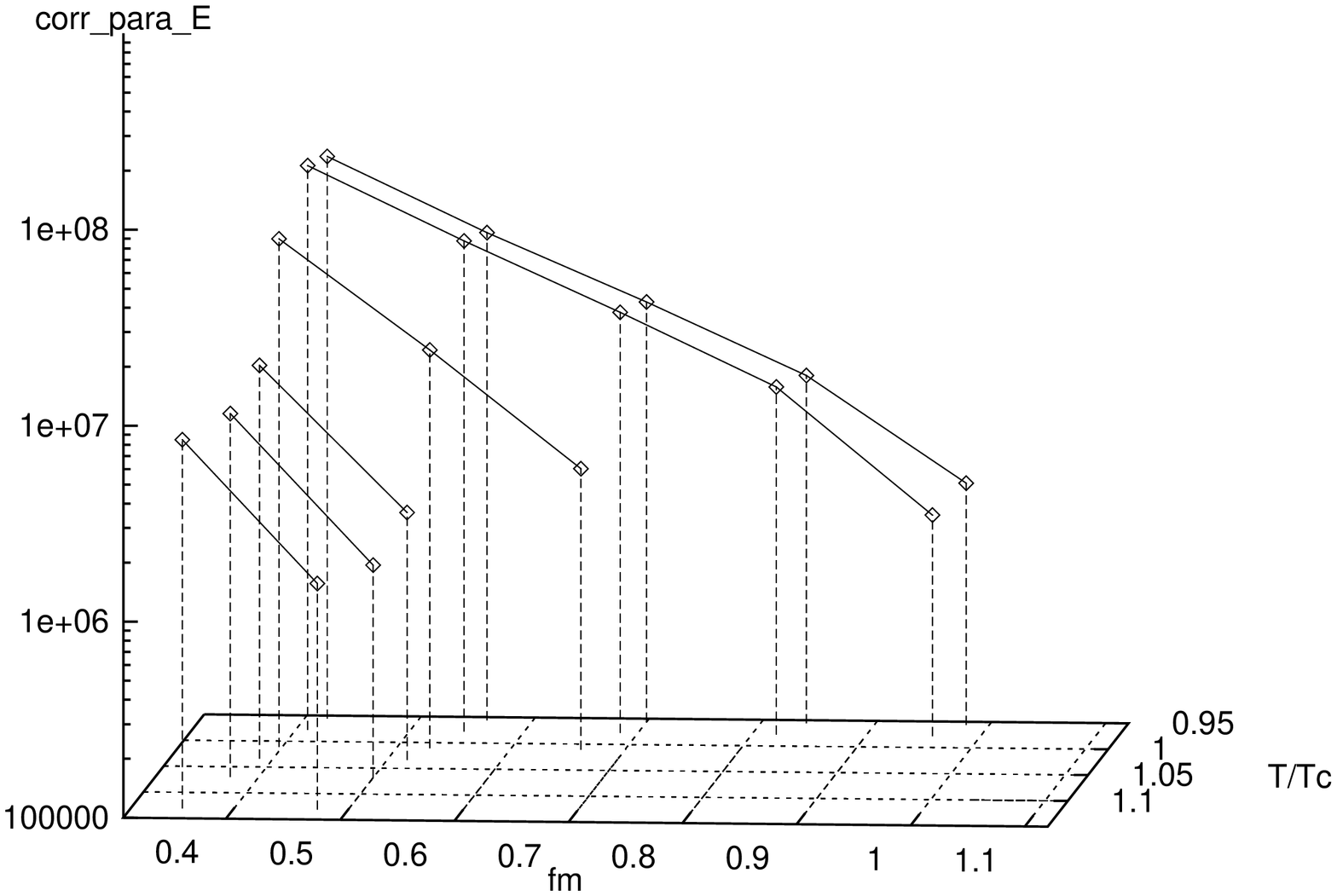}}
{\bf Fig.2} The electric longitudinal correlator versus distance,
for different values of $T/T_c$.
\end{minipage}
\par\noindent
\par\noindent
\begin{minipage}{0.9\linewidth}
\epsfxsize0.8\linewidth
\rotatebox{270}{
\epsfbox{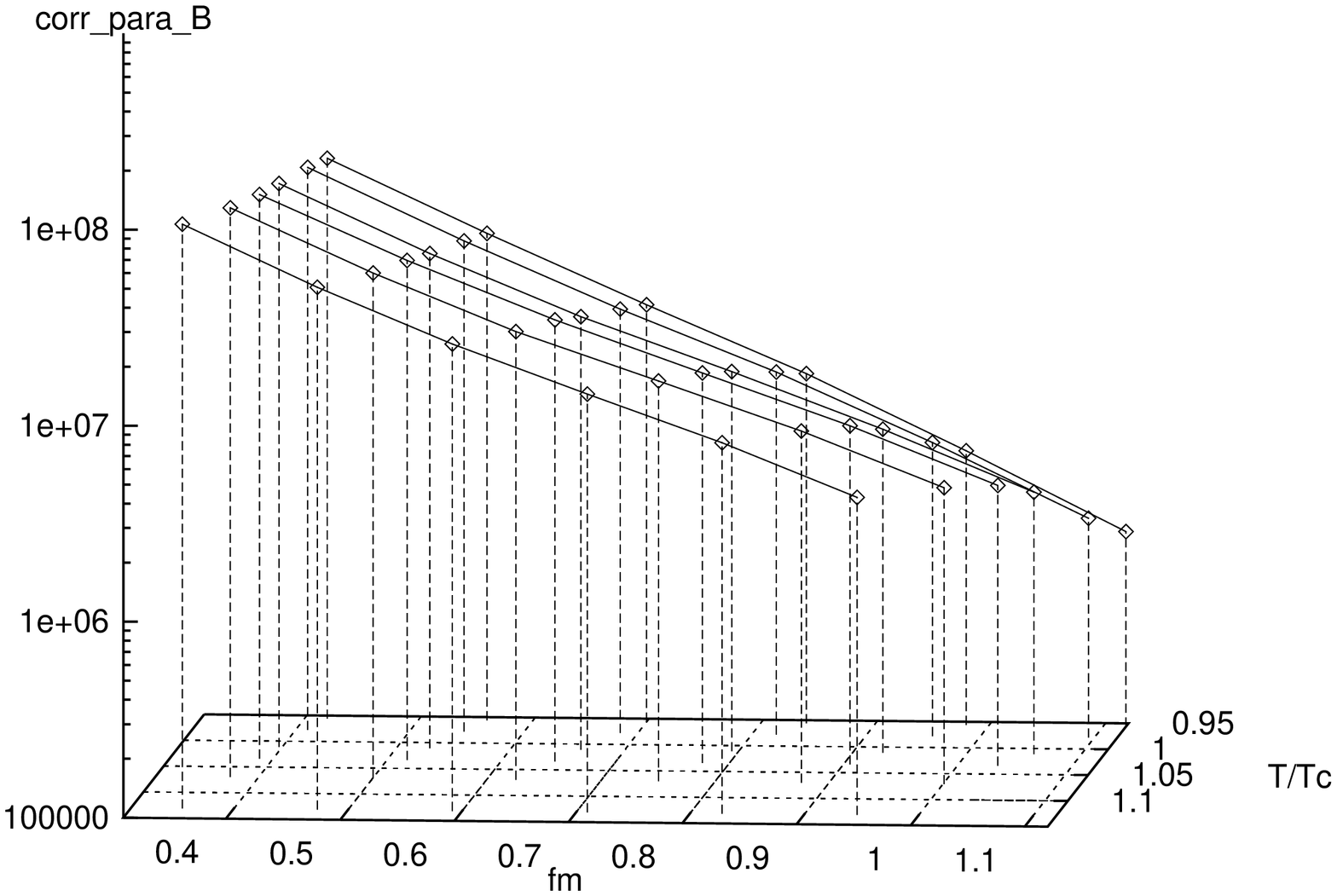}}
{\bf Fig.3} The magnetic longitudinal correlator versus distance,
for different values of $T/T_c$.
\end{minipage}
\par\noindent
The scalar quark correlator
\begin{equation}
S_0(x) = - \langle0| T(\bar\psi(x) S \psi(0))|0\rangle
\label{eq:20}\end{equation}
has recently been determined in full QCD and in quenched QCD\cite{19}.

A comparison has been made between the determinations in full QCD, at  given
values of the quark masses $am_q=0.01$, $am_q=0.02$
and of the lattice spacing $a$, and in quenched QCD at the same values of these
physical parameters. No difference has been found, within errors, indicating
that quark loops do not affect appreciably the quark correlator.

A sensible
parametrization for the lattice regulator $S_0^L$ is
\begin{equation}
S_0^L(x) = a^3 A_0 \exp(-x/\lambda_f) + \frac{B_0
a^3}{x^2}\label{eq:21}\end{equation}
Simulations have been performed\cite{19} at $a m_q = 0.01$ in full QCD at
$\beta = 5.35$ and quenched QCD at $\beta = 6.0$, which correspond both to a
lattice spacing $a\simeq 0.10$~fm. No appreciable difference is found and in
both cases (fig.4)
\par\noindent
\begin{minipage}{0.9\linewidth}
\epsfxsize0.8\linewidth
\rotatebox{270}{
\epsfbox{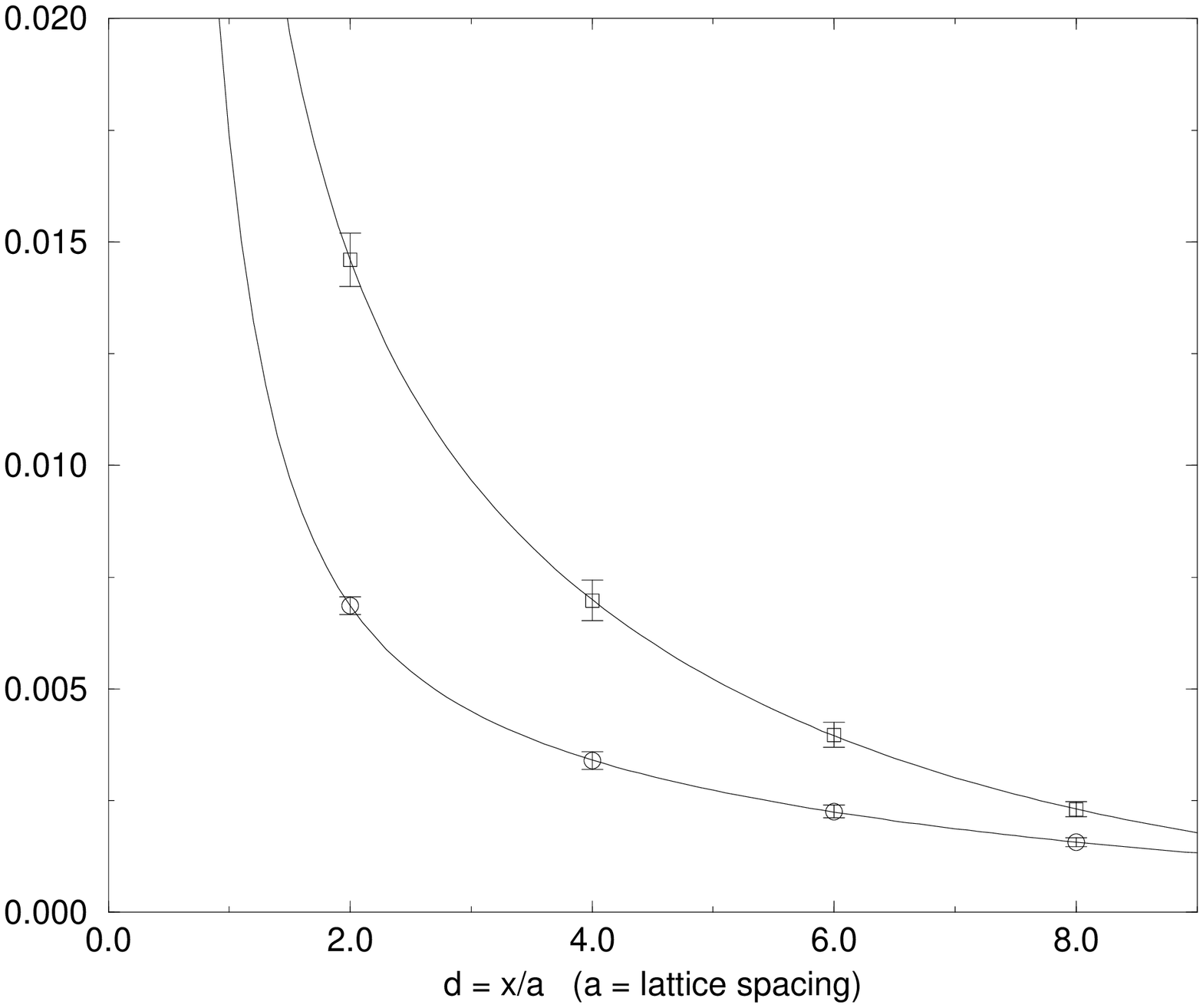}}
{\bf Fig.4} $S_0^L(x)$ versus $x$. The curve is the best fit
to eq.(23).
\end{minipage}
\par\noindent
\begin{equation}
\frac{a}{\lambda_f} = 0.16\pm 0.04\qquad a m_{\pi} =
0.26\pm0.01\label{eq:22}\end{equation}
A similar determination at $a m_q = 0.02$ $\beta = 5.35$ full QCD, $\beta =
5.91$ quenched, where $a\simeq 0.12$~fm, give indistinguishable results:
\begin{equation}
\frac{a}{\lambda_f} = 0.26\pm 0.04\qquad a m_{\pi} =
0.37\pm0.01\label{eq:23}\end{equation}
Putting the two determinations together gives
\begin{equation}
\lambda_f m_\pi = 1.5\pm 0.3\label{eq:24}\end{equation}
The typical correlation length is now $\sim1/m_\pi$ in agreement with the
approach of ref.\cite{3}.

A determination of $A_0$ and a study of its relation to the quark condensate
$\langle\bar\psi\psi\rangle$ is on the way.

\section{Discussion}
The typical correlation length is rather small for gluon correlators:
$\lambda =
0.16$~fm for quenched $SU(2)$, $0.22$~fm for quenched $SU(3)$, $0.32$~fm for
full
QCD with 4 flavours. It is bigger for fermion correlators where
\[ \lambda_f \simeq (1.5\pm 0.3)m_\pi^{-1}\]
Condensates can be determined, despite the presence of renormalons, by the
same
philosophy used for SVZ sum rules. $G_2 = (0.022\pm0.006)$~GeV$^4$ in full QCD
is consistent with the determination by sum rules. It is an order of magnitude
bigger in quenched QCD, $G_2 = (0.14\pm0.02)$~GeV$^4$.

The behaviour of correlators at the deconfining transition is consistent with
expectations.

Our determinations are useful to phenomenology and to test models of QCD
vacuum.
\section*{Acknowledgements}
Partially supported by EC TMR Program
ERBFMRX-CT97-0122, and by MURST, project: ``Fisica Teorica delle
Interazioni Fondamentali''.


\section*{References}
\begin{thebibliography}{99}
\bibitem{1}H.G. Dosch, \Journal{\PLB}{190}{177}{1987}.
\bibitem{2}H.G. Dosch, Yu. A. Simonov, \Journal{\PLB}{205}{339}{1988}.
\bibitem{3}M.A. Shifman, A.I. Vainshtein, V.I. Zakharov,
\Journal{\NPB}{147}{385,448,519}{1979}.
\bibitem{4}H. Leutwyler, \Journal{\PLB}{98}{447}{1981}.
\bibitem{5}M.B. Voloshin, \Journal{\NPB}{154}{365}{1979}.
\bibitem{6}M. Campostrini, A. Di Giacomo, \v{S}.~Olejn\'{\i}k,
\Journal{\ZPC}{31}{5}{1986}.
\bibitem{7}M. Campostrini, A. Di Giacomo, G. Mussardo,
\Journal{\ZPC}{25}{173}{1984}.
\bibitem{8}S.V. Mikhailov, A.V. Radjushkin,
\Journal{\JPL}{43}{713}{1986}.
\bibitem{9}M. Campostrini, A. Di Giacomo, M. Maggiore, H. Panagopoulos,
E. Vicari, \Journal{\PLB}{225}{403}{1989}.
\bibitem{10}A. Di Giacomo, H. Panagopoulos,
\Journal{\PLB}{285}{133}{1992}.
\bibitem{11}A. Di Giacomo, E. Meggiolaro, H. Panagopoulos,
\Journal{\NPB}{483}{371}{1997}.
\bibitem{12}M. D'Elia, A. Di Giacomo,  E. Meggiolaro,
\Journal{\PLB}{408}{315}{1997}.
\bibitem{13}A.H. M\"uller,
\Journal{\NPB}{250}{327}{1985}.
\bibitem{14}V.A. Novikov, M.A. Shifman, A.I. Vainshtein, V.I. Zakharov,
\Journal{\NPB}{198}{301}{1981}.
\bibitem{15}S. Narison,
\Journal{\PLB}{387}{162}{1996}.
\bibitem{16}A. Di Giacomo, G.C. Rossi,
\Journal{\PLB}{100}{481}{1981}.
\bibitem{17}A. Di Giacomo, G. Paffuti,
\Journal{\PLB}{108}{327}{1982}.
\bibitem{18}M. Campostrini, A. Di Giacomo, Y. G\"und\"uc,
\Journal{\PLB}{225}{393}{1983}.
\bibitem{19}M. D'Elia, A. Di Giacomo,  E. Meggiolaro,
in preparation.
\end{thebibliography}
\end{document}